\documentclass[english,showpacs,twocolumn,superscriptaddress]{revtex4-1}
\usepackage{mathptmx}
\usepackage[T1]{fontenc}
\usepackage[latin9]{inputenc}
\usepackage{color}
\usepackage{amsmath}
\usepackage{amssymb}
\usepackage{hyperref}
\usepackage{graphics}
\usepackage{graphicx}
\usepackage{epsfig}
\usepackage{bm}
\usepackage{epstopdf}
\usepackage{mathtools}
\usepackage{siunitx}
\usepackage{float}
\graphicspath{{Figures/}}

\usepackage{lineno}

\usepackage{xmpmulti}

\usepackage{babel}

\begin{document}

%\setpagewiselinenumbers
%\modulolinenumbers[1]
%\linenumbers

\title{Effect of interparticle interaction on motility induced phase separation of self-propelled inertial disks}
\author{Soumen De Karmakar}
\email{soumendekarmakar@gmail.com}
\author{Rajaraman Ganesh}
\email{ganesh@ipr.res.in}
\affiliation{Institute for Plasma Research, Bhat, Gandhinagar 382428, Gujarat, India}
\affiliation{Homi Bhabha National Institute, Training School Complex, Anushaktinagar, Mumbai 400094, India}

\date{\today}

%**************************************************************
%ABSTRACT
%**************************************************************
\begin{abstract}
  Phase diagram of the phenomenon of motility induced phase separation (MIPS) for a collection of self-propelled interacting disks is explored using Langevin dynamics simulation with particular emphasis on disk wall softness and the range of interaction amongst disks. We bring out important changes in the MIPS phase diagram both due to softness and inertia of the disks. Specifically, we show that overdamped softer disks phase separate while MIPS becomes possible only for harder disks in the inertial limit.  Unlike most of the earlier studies on MIPS which consider hard-core disks, our findings may be directly applicable to soft active matter for a range of biological systems. 
\end{abstract}

%%%%%%%%%%%%%%%%%%%%%%%%%%%%%%%%
\maketitle
%%%%%%%%%%%%%%%%%%%%%%%%%%%%%%%%

%-----------------------------------------------------------------------
%-----------------------------------------------------------------------
%Introduction
%-----------------------------------------------------------------------
Motility induced phase separation (MIPS) is one of the fundamental properties of self-propelled or active particles due to their persistent motion\cite{ref1, ref2}. Fundamental criteria behind MIPS is the blockage of free propulsion during collision. Unlike passive particles, due to their persistent motion, two self-propelled particles tend to remain together upon collision, until the self-propulsion direction of one of the particle turns away from the other due to inherent fluctuation. Above a minimum value of particle density and persistence time, MIPS is observed in several experimental systems\cite{ref3, ref4, ref5} as well as in numerical simulation\cite{ref1, ref6, ref7, ref8}. In some recent studies, the complete phase diagram in the density-persistence space for two dimensional self-propelled disks have been obtained in numerical simulation\cite{ref7, ref9, ref10}. Properties of the different phases have been studied to some extent\cite{ref6, ref11}. In some recent work, various types of alignment mechanism of the motile particles on MIPS have been studied\cite{ref12, ref13}. Some of the alignment mechanism favor MIPS\cite{ref14, ref15, ref16}, the others suppress MIPS\cite{ref17, ref18}.

Most of the studies, so far, have considered the overdamped limit of the self-propelled particles. Inertia brings in important differences not only in the dynamical properties, but also in the structural and steady state properties such as active temperature and pressure\cite{ref19, ref20, ref21, ref22, ref23}. A recent numerical study\cite{ref24} of inertial self-propelled disks on MIPS found that MIPS vanishes at large inertia. Moreover, unlike the overdamped self-propelled disks, where the low and high density phases possess the same temperature, different temperature was obtained in the two distinct phases in the inertial limit. The low density phase remains at the higher temperature and the high density phase acquires a low temperature.

Despite the important role of inter-particle interaction, most of the studies of MIPS focus on either excluded volume interaction or sufficiently hard-core and short ranged repulsion, e.g., Weeks-Chandler-Andersen (WCA)\cite{ref12} type interaction, at the walls of the self-propelled particles. Redner et at.\cite{ref25} demonstrated that a small attractive component in the interaction potential modify the phases of the self-propelled disks, from active gel-like state to motility induced clusters. Yan et al.\cite{ref26} demonstrated various collective states by changing the inter-particle interactions between the Janus spheres. There are numerous examples of biological active matter ranging from sub-cellular elements to tissues\cite{ref27}, whose walls are sufficiently soft. Hence, the interaction potential, which models those soft active materials, should have much softer repulsion at the boundary of the particles compared to WCA type potential. Furthermore, several propulsion mechanism produces the inter-particle interaction ranging several particle diameters\cite{ref26, ref28}. 

In this study, we incorporate finite size self-propelled disks in two dimensions, interacting with the other disks through modified Yukawa potential (see Fig.~\textcolor{blue}{\ref{fig:fig1}} and the corresponding expression given below). Choice of our interaction potential enable us to control the softness or the stiffness at the walls of the disks and the range of the interaction by tuning the stiffness and the strength parameters of the interaction potential. Hence, we are able to study interacting self-propelled disks, from much softer to a sufficiently stiffer ones. Moreover, we have the control over the range of the interaction, from a short range to relatively long range one. Furthermore, we have considered finite inertia of the self-propelled disks. Controlling the inertial parameter, we are able to move from a low inertia to sufficiently large inertial domain. We demonstrate that softness of the interaction and the inertia modify the phase diagram of our system in several important ways.

%-----------------------------------------------------------------------
%-----------------------------------------------------------------------
%Description of the Numerical Model
%-----------------------------------------------------------------------
We perform the Langevin dynamics simulation of $N$ self-propelled inertial disks (SPIDs) of uniform mass $m$, moment of inertia $I$, and diameter $\sigma$. Due to their inherent motility, a self-propulsion direction $\boldsymbol{n}_i = (\cos\theta_i, \sin\theta_i)$ is associated to each disk. Dynamics of the center of mass velocities and the orientations $\{ \textbf{v}_i, \theta_i \}$ are governed by $N$ translational and rotational Langevin equations,
  %
  %\begin{equation}
  \begin{align}  
    m\dot{\textbf{v}}_i + \gamma \textbf{v}_i & = \textbf{F}_i + \textbf{F}^a_i + \sqrt{2 \gamma^2 D} \boldsymbol{\xi}_i \label{eqn:eqn1},\\
    I\ddot{\theta}_i + \gamma_r \dot{\theta}_i & = \sqrt{2 \gamma_r^2 D_r} \zeta_i\label{eqn:eqn2},
  \end{align}
  %\end{equation}
  %
respectively. The suffix $r$ denotes rotational parameters. $\gamma$, $\gamma_r$ are dissipation coefficients. We assume diffusion coefficients, $D$ and $D_r$, obey fluctuation-dissipation relation, and they are coupled as $D = \sigma^2 D_r$\cite{ref29}. $\boldsymbol{\xi}_i$ and $\zeta_i$ are the white Gaussian noises. The disks self-propel with a constant propulsion speed $\textup{v}_0$ along the propulsion direction $\boldsymbol{n}_i(t)$. The associated self-propulsion force is $\textbf{F}_i^a (t) = \gamma \textup{v}_0 \boldsymbol{n}_i(t)$. A disk interacts with the other disks with purely repulsive pairwise force ($\textbf{F}_{ij}$) such that the total conservative potential energy of the system is $U = \displaystyle{\sum_{i<j}} V_0 \frac{e^{-(r_{ij} - \sigma) / \lambda}}{r_{ij}}$. By adjusting the interaction parameters $\lambda$ and $V_0$, the softness and the range of interaction of the SPIDs is varied (see Fig.~\textcolor{blue}{\ref{fig:fig1}}). 
We consider $\sigma$, $1/D_r$, and the background thermal energy $k_B T$ as the unit of normalization for length, time, and energy, respectively. $N$ normalized dynamical equations for the self-propelled inertial disks are:
  %
  %\begin{equation}
  \begin{align}  
    M\dot{\textbf{v}}_i + \textbf{v}_i & = -\boldsymbol{\nabla}_i \Gamma \sum_{i<j} \frac{e^{-\kappa (r_{ij} - 1)}}{r_{ij}} + \textup{P}_e \textbf{n}_i + \sqrt{2} \boldsymbol{\xi}_i \label{eqn:eqn4},\\
    J \ddot{\theta}_i + \dot{\theta}_i & = \sqrt{2} \zeta_i \label{eqn:eqn5}.
  \end{align}
  %\end{equation}
  %
  $\boldsymbol{\nabla}_i$ is the gradient operator at the location of the $i$'th particle $\boldsymbol{r}_i$. Inertial parameter $M = \frac{m / \gamma}{1 / D_r}$ is the ratio of inertial time scale ($m / \gamma$) to the persistence time scale ($\tau_p = 1 / D_r$). Rotational inertial parameter $J = \frac{I / \gamma_r}{1 / D_r}$ is the ratio of rotational inertial time scale ($I / \gamma_r$) to the persistence time scale. Peclet number $\textup{P}_e = \textup{v}_0 \tau_p / \sigma$ is the ratio of persistence length ($l_p = \textup{v}_0 \tau_p$) to the the diameter of the self-propelled disks. Reduced interaction parameters are $\Gamma = \frac{V_0 / \sigma}{k_B T}$, $\kappa = \sigma / \lambda$. All subsequent results are provided in the reduced units.

  We plot the interaction potential for $\Gamma = 25$ in Fig.~\textcolor{blue}{\ref{fig:fig1}(a)}, and $\kappa = 14$ in Fig.~\textcolor{blue}{\ref{fig:fig1}(b)} with the solid lines. For comparison, WCA interaction with unit strength is plotted with the dashed lines.
%
%Fig 1
\begin{figure}[ht]
  \includegraphics[width=4.28cm, height=4.0cm]{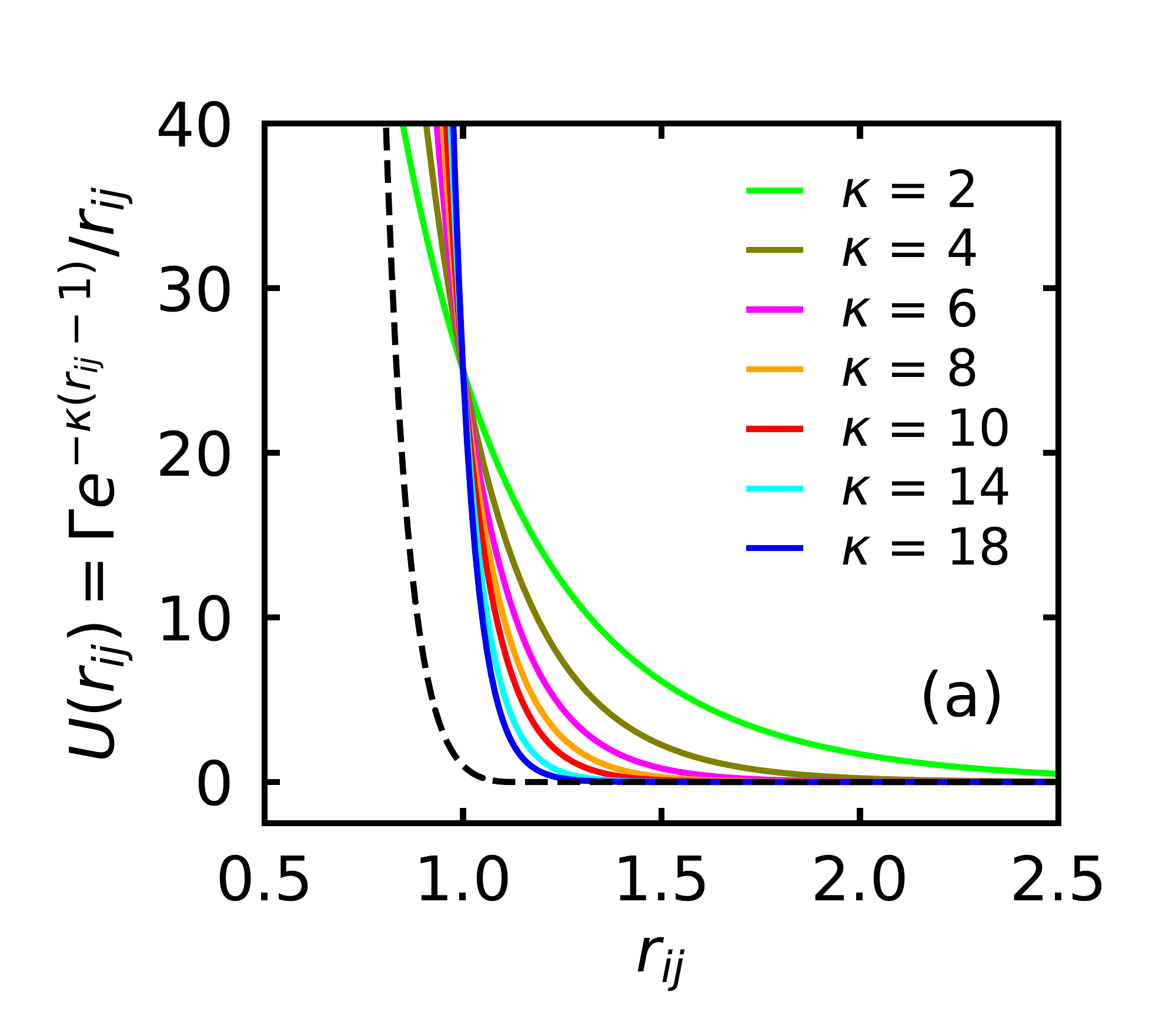}
  \includegraphics[width=4.28cm, height=4.0cm]{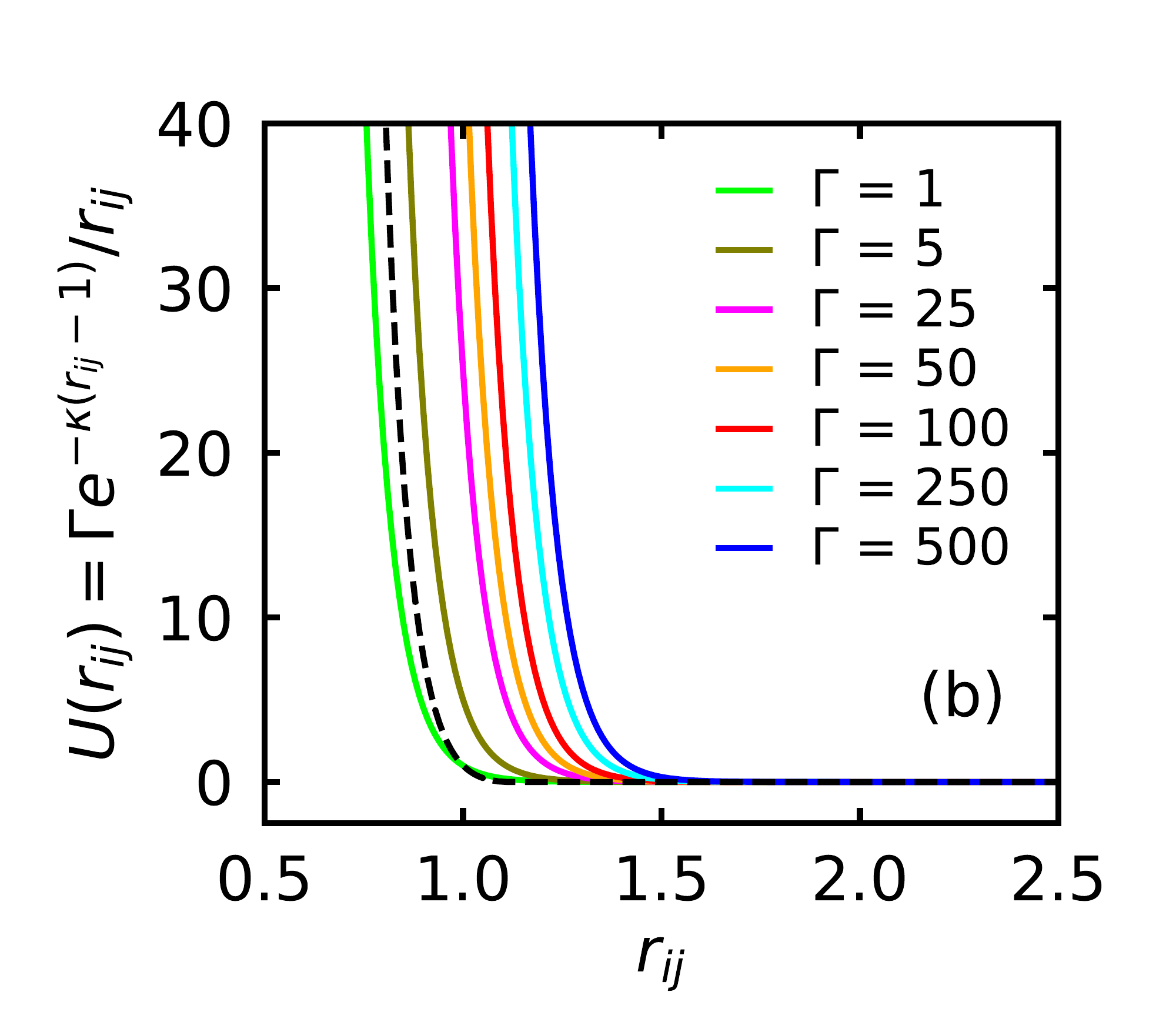}
  \caption{\label{fig:fig1} Plot of pairwise interaction ($U(r_{ij})$) for (a) $\Gamma = 25.0$ and several values of $\kappa$, (b) $\kappa = 14.0$ and several values of $\Gamma$, shown in the legend. WCA potential\cite{ref12, ref24} with unit strength is shown by the dashed lines.}
\end{figure}
The softness of the interaction matches to that of WCA interaction softness for $\kappa \approx 18$. For relatively low values of $\kappa$, the disks become soft. We call $\kappa$ as the softness or the stiffness parameter. Moreover, with decreasing $\kappa$, the range of interaction increases. For a fixed stiffness, the strength of the potential $\Gamma$ dictates the effective size of the disks.

  We consider $N = 48400$ SPIDs in a two-dimensional rectangular simulation box of dimensional ratio $L_x / L_y = 2 / \sqrt{3}$ and perform Langevin dynamics simulation with the upgraded GPU based Molecular Dynamics solver MPMD\cite{ref20}. Packing fraction, $\phi = N \pi / 4 L_x L_y$, and Peclet number are fixed at $\phi = 0.5$ and $\textup{P}_e = 75$, respectively, unless, otherwise, specified. We have checked that the considered values of $\phi$ and $\textup{P}_e$ are sufficient to produce MIPS in the overdamped limit and for sufficiently stiff interaction parameter ($\kappa$), tending towards WCA interaction\cite{ref24}. We consider doubly periodic boundary conditions. Interaction cut-off is set at $r_c = 8$. Integration time step is fixed at $\Delta t = 10^{-4}$, which gives good energy conservation in the steady state for the whole parameter space that we have considered.

\begin{figure*}[ht]
  \includegraphics[width=5.9cm, height=4.5cm]{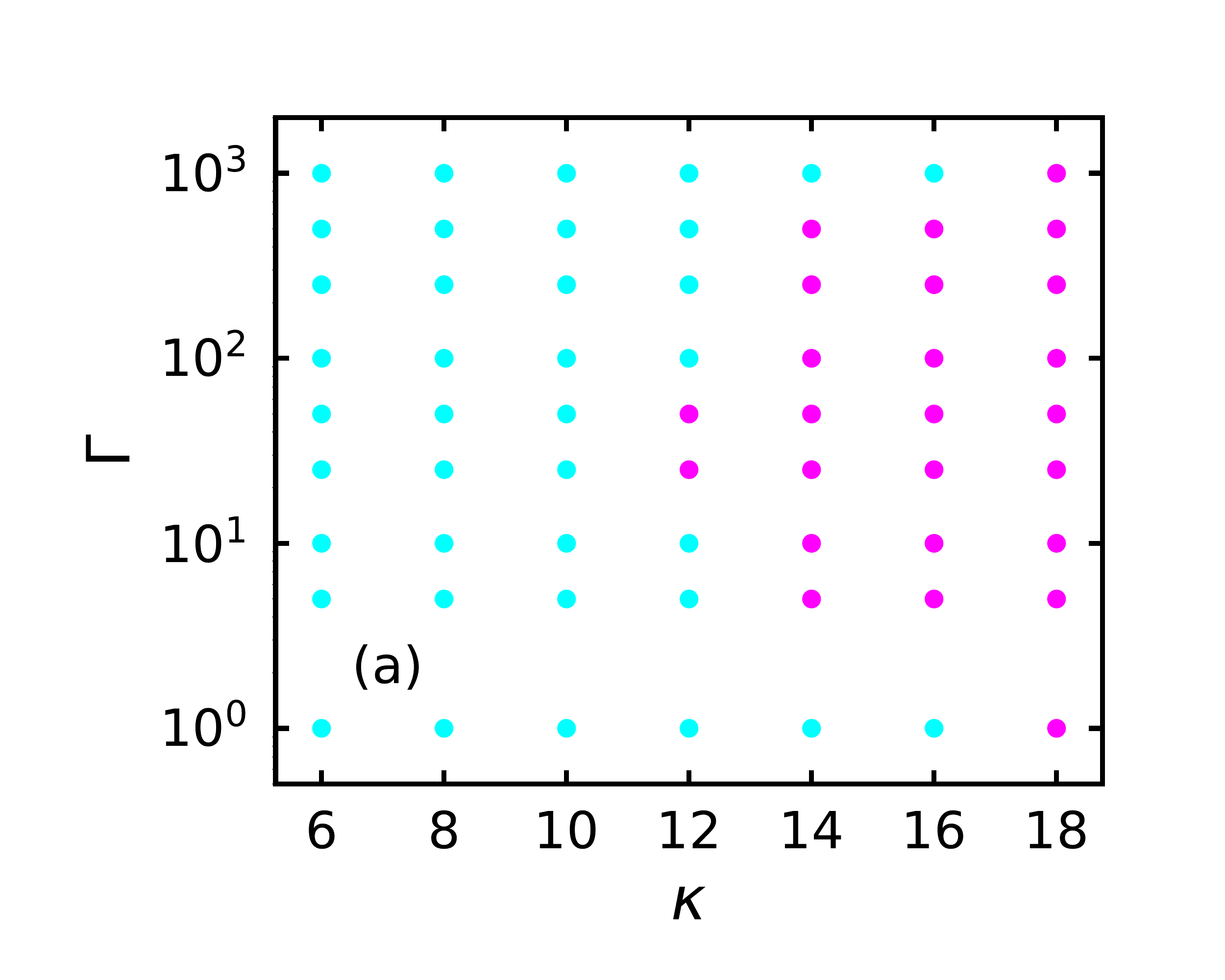}
  \includegraphics[width=5.9cm, height=4.5cm]{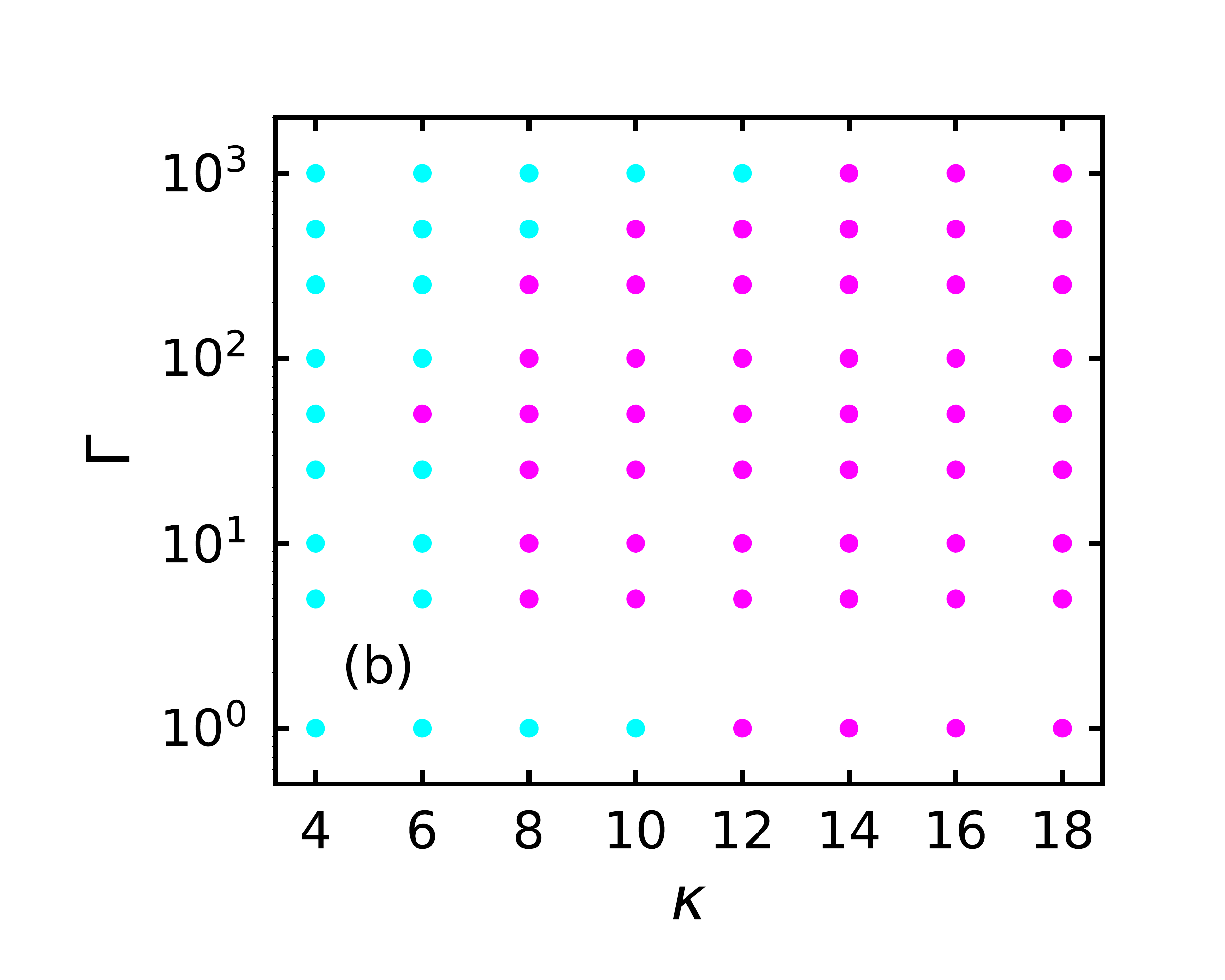}
  \includegraphics[width=5.9cm, height=4.5cm]{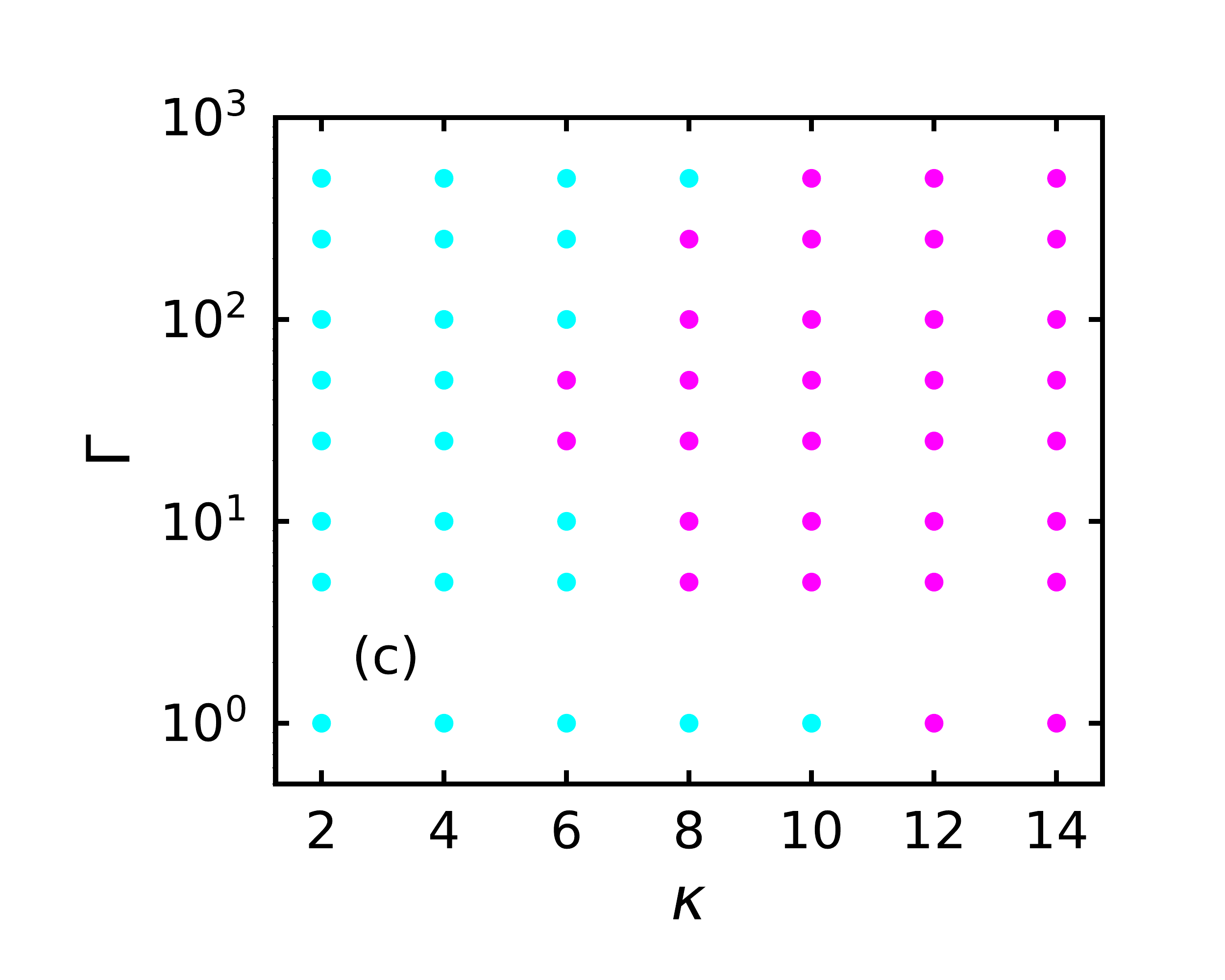}
  \caption{\label{fig:fig2} Phase diagram in the $\kappa-\Gamma$ space for (a) $M = 0.05$, (b) $M = 0.01$, and (c) $M = 0.005$. The value of $J$ is fixed at $J= 0.01$. The magenta and blue dots denote the MIPS and homogeneous phases, respectively.}
\end{figure*}

\begin{figure*}[ht]
  \includegraphics[width=5.9cm, height=4.5cm]{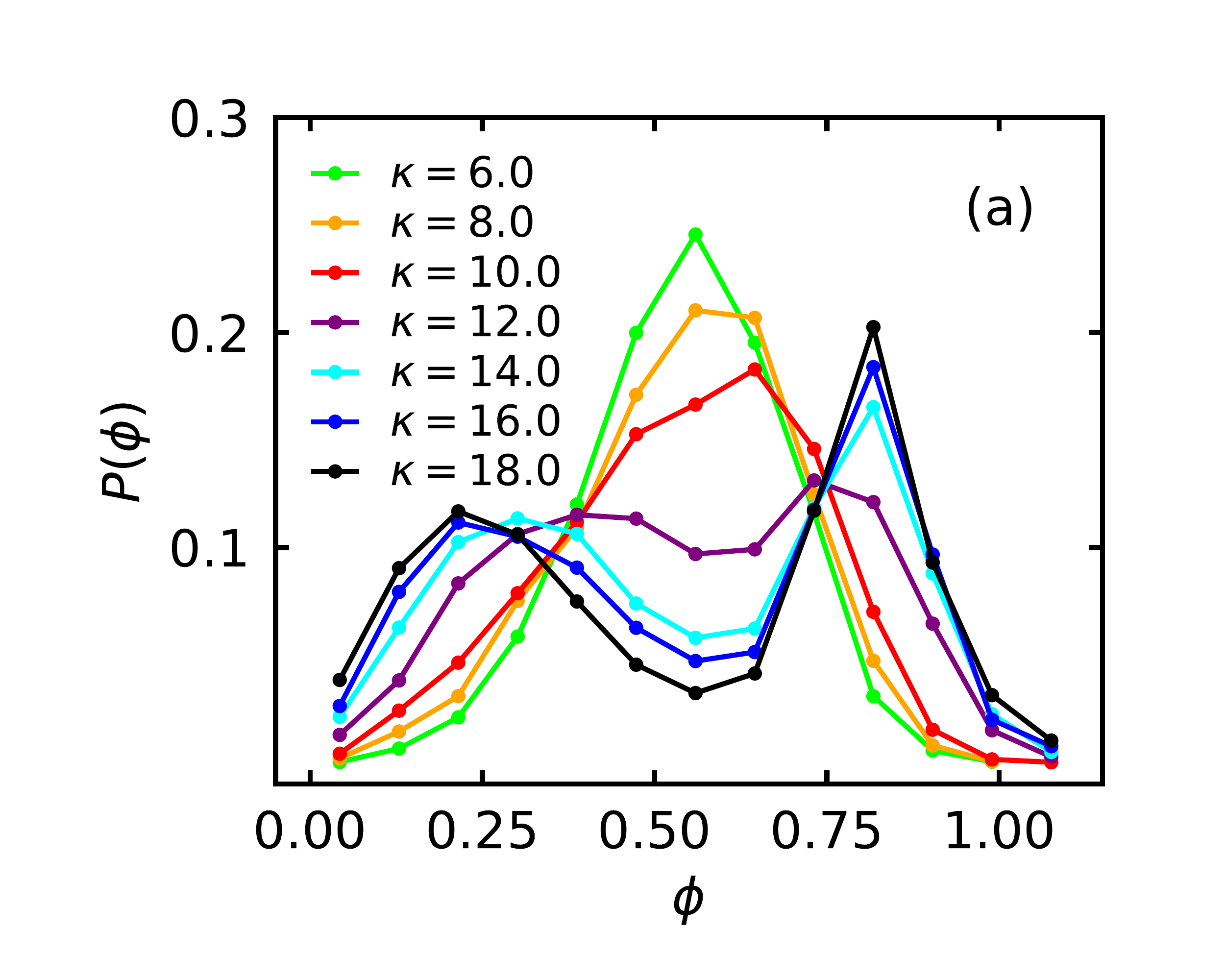}
  \includegraphics[width=5.9cm, height=4.5cm]{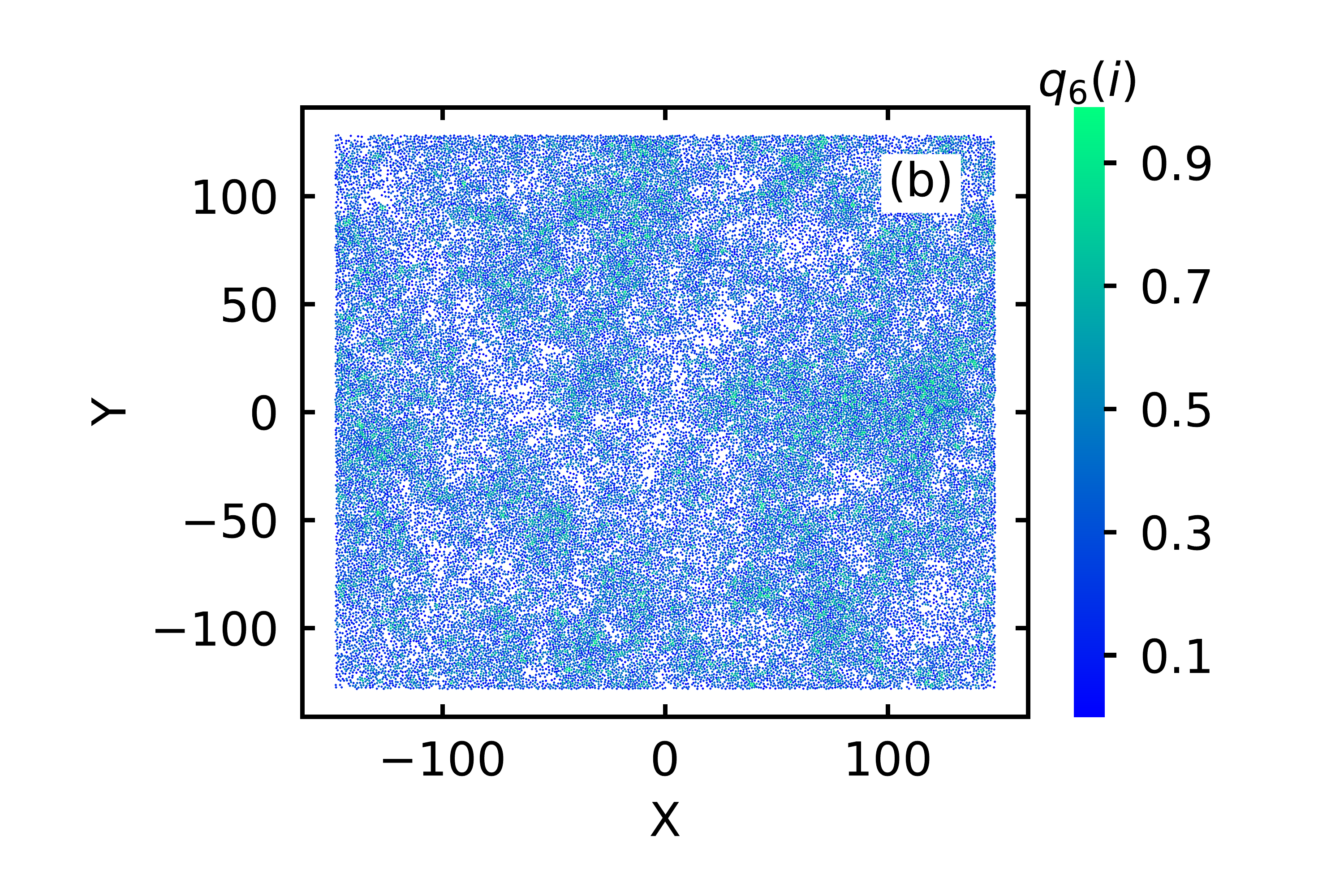}
  \includegraphics[width=5.9cm, height=4.5cm]{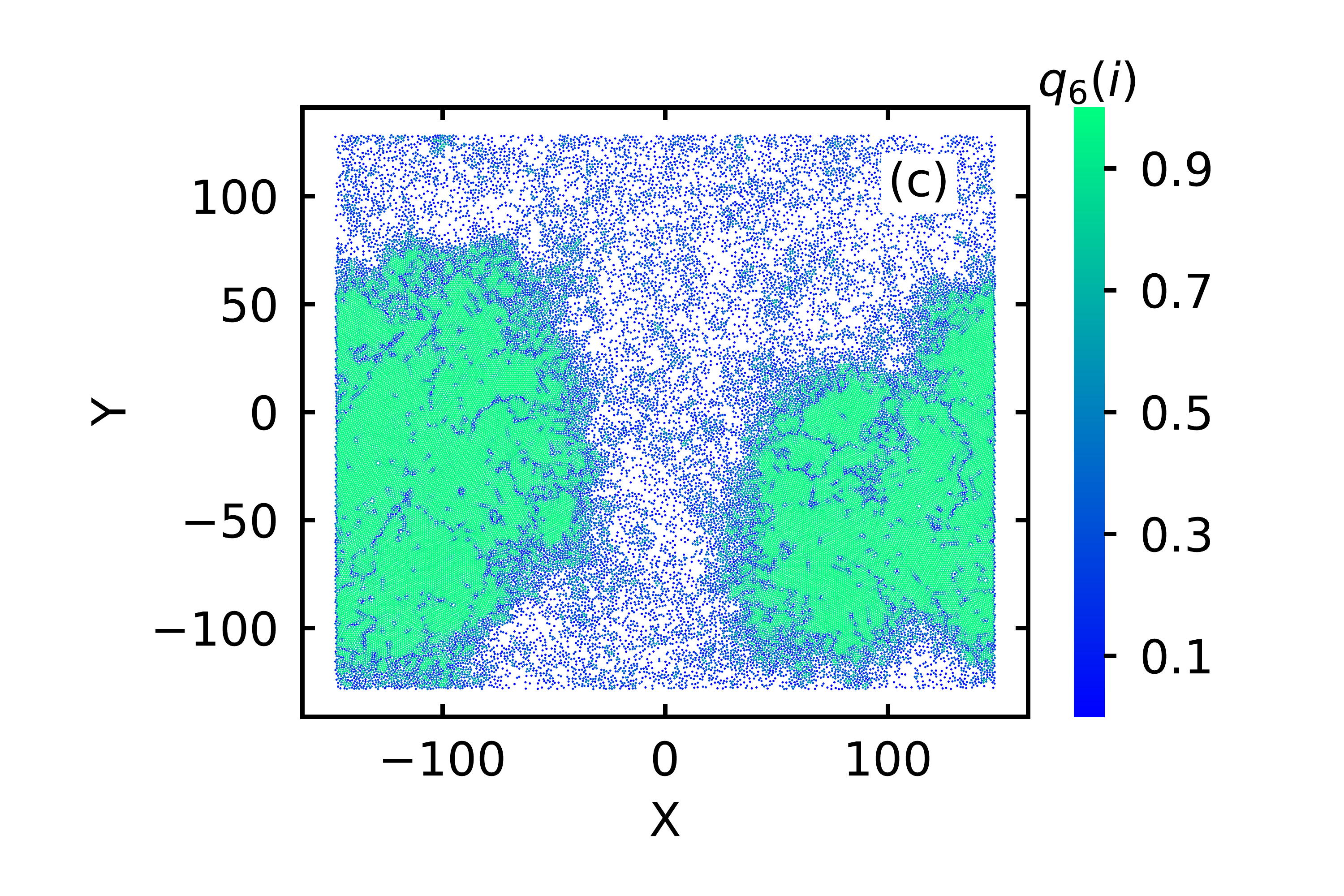}
  \caption{\label{fig:fig3} (a) Distribution of local area fraction for various values of $\kappa$, shown in the legend, and fixed $\Gamma = 25$. Configuration plots of the homogeneous (b), and phase separated (c) systems, corresponding to $\kappa = 6$ (soft) and $\kappa = 16$ (stiff), respectively, of (a). Local orientational order at the location of the disks are shown in the color bars (see the main text for the definition of $q_6(i)$). The inertial parameter values are fixed at $M = 0.05$, and $J = 0.01$.}
\end{figure*}

\begin{figure}[ht]
  \includegraphics[width=8.5cm, height=6.5cm]{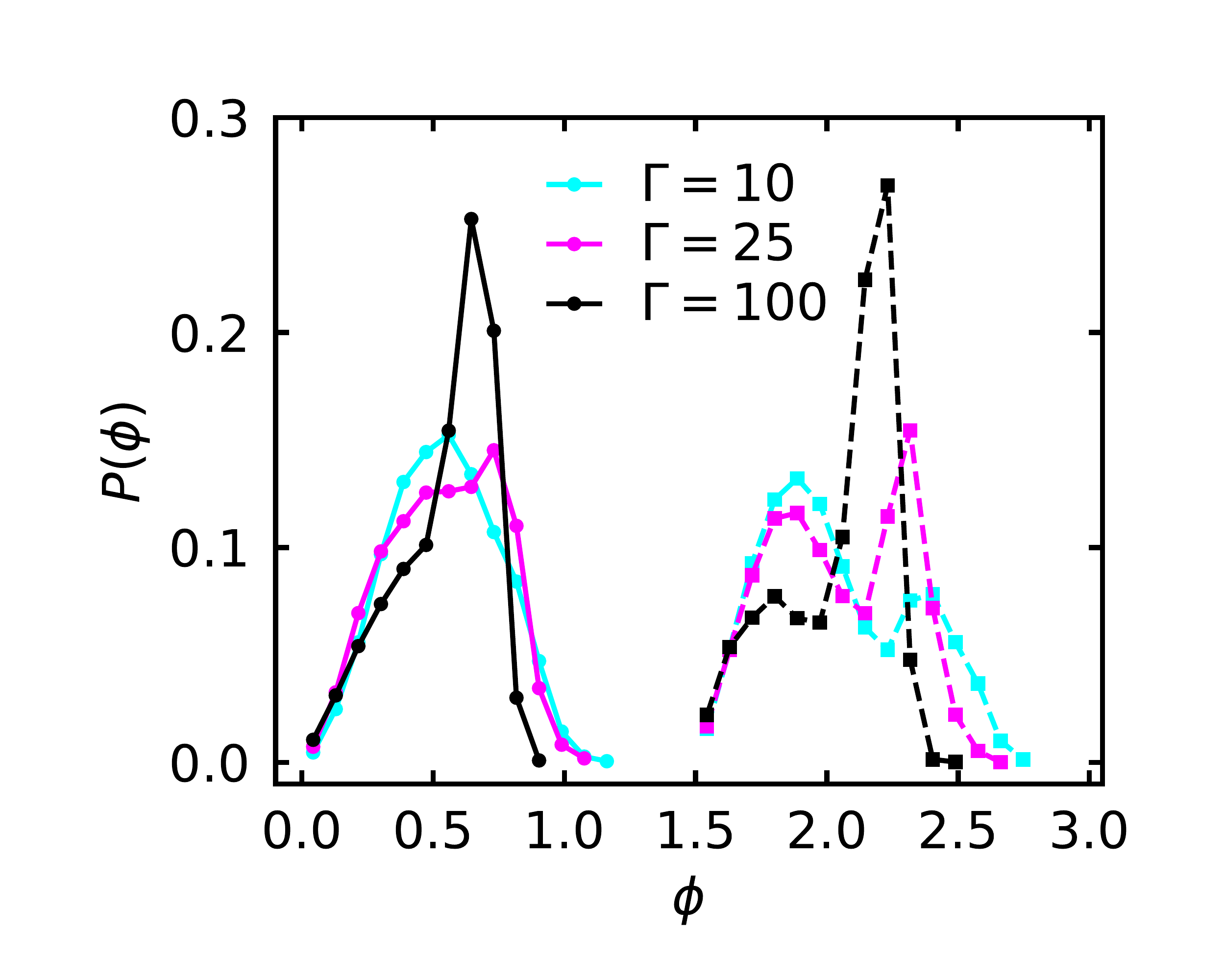}
  \caption{\label{fig:fig4} Distribution of local area fraction for $J = 0.0001$ (solid lines) and $J = 0.1$ (dashed lines) for three different values of $\Gamma$, shown in the legend. Inertial and softness parameters are fixed at $M = 0.05$, and $\kappa = 12$, respectively. Dashed lines are shifted along the $\phi$ axis by $1.5$ units, for better visibility.}
\end{figure}

\begin{figure}[ht]
  \includegraphics[width=7.0cm, height=8.5cm]{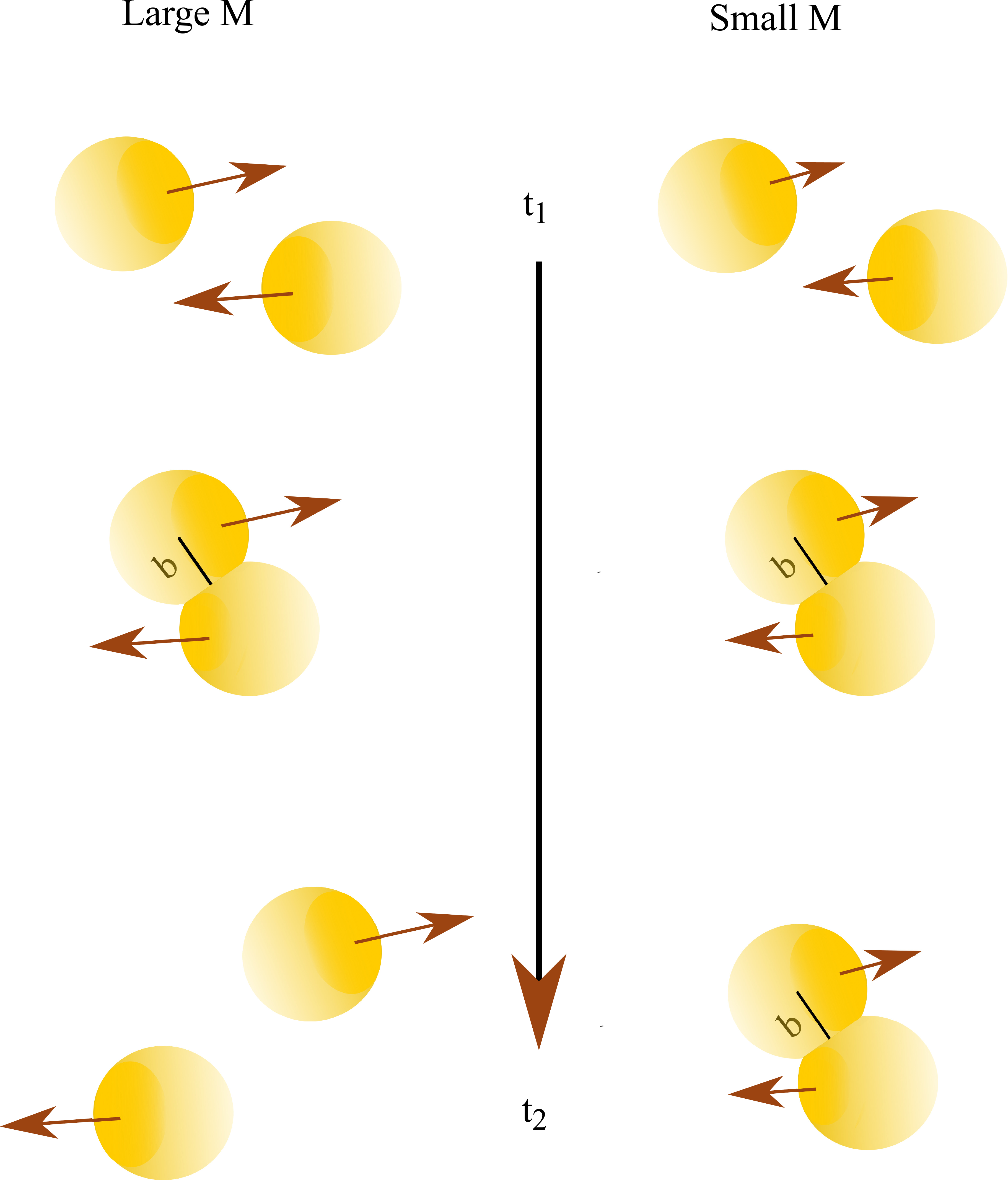}
  \caption{\label{fig:fig5} Diagram depicting the collision process of two soft SPIDs with large (left column) and small (right column) value of inertial parameter. Due to the high kinetic energy, when two soft inertial disks (left column) with large impact parameter $b$ collide, the shape is deformed and they move apart with minor change in the propulsion directions. However, when two soft disks with small inertial value $M$ collide with a large impact parameter $b$, the shape is deformed but they stay together, due to their low kinetic energy, during their persistence time. Hence, provide local nucleation sites for cluster formation.}
\end{figure}

  In Fig.~\textcolor{blue}{\ref{fig:fig2}}, we plot the phase diagram of our system of inertial SPIDs in the $\kappa-\Gamma$ space. The rotational inertial parameter $J$ is fixed at $J = 0.01$. $M$ is varied from $M = 0.05$ (Fig.~\textcolor{blue}{\ref{fig:fig3}(a)}) to $M = 0.005$ (Fig.~\textcolor{blue}{\ref{fig:fig3}(b)}) through $M = 0.01$ (Fig.~\textcolor{blue}{\ref{fig:fig3}(c)}) from left to right.
%
%Fig 2
Blue and magenta dots indicate the homogeneous and phase separated states, respectively. We observe a parabola shaped phase boundary separating the MIPS and homogeneous regions over an order of magnitude variation in $M$. We measure the phase separation from the two distinct peaks in the area fraction distribution across the mean value, which is fixed at $\phi = 0.5$. In Fig.~\textcolor{blue}{\ref{fig:fig3}(a)}, we plot the distribution of local area fraction for $\Gamma = 25$, and several values of $\kappa$, shown in the legend for $M = 0.05$, $J = 0.01$. The single peaked distribution around $\phi = 0.5$ becomes doubly peaked for $\kappa \geq 12$. Unlike passive disks, the distribution is stretched along the $\phi$ axis due to self-propulsion of the disks (not shown). Self-propulsion makes the disks explore the extreme density ranges, overcoming the effective size (see Fig.~\textcolor{blue}{\ref{fig:fig5}(a)}) set by the interaction potential. In Figs.~\textcolor{blue}{\ref{fig:fig3}(b), (c)}, we plot the configuration of the system in the homogeneous and the phase separated phases at $\kappa = 6$ and $\kappa = 16$, respectively, corresponding to Fig.~\textcolor{blue}{\ref{fig:fig3}(a)}. The color bars denote the local orientational order\cite{ref30, ref31} of the disks, given by
\begin{equation}
  q_6(i) = \frac{1}{6} \sum_{j=1}^{6} e^{i6\theta_{ij}}.
  \label{eqn:eqn5}
\end{equation}  
$\theta_{ij}$ is the angle between the $i$ and $j$ disks with respect to some arbitrary fixed axis. The ordered region in Fig.~\textcolor{blue}{\ref{fig:fig3}(c)}, which corresponds to the high density peak in the local density distribution, is surrounded by low density disordered region. The homogeneous phase in Fig.~\textcolor{blue}{\ref{fig:fig3}(b)} has no order. Near the phase boundary, in the homogeneous side of the phase diagram, we observe few relatively high density scattered regions, which do not grow in the simulation time to produce two distinct peaks in the distribution of local area fraction (not shown). Rather, those regions vanish and emerge at some other location with time, and maintain an overall dynamic homogeneous phase.
%Fig 3
We plot the distribution of the local area fraction for $J = 0.0001$ and $J = 0.1$ by the solid and dashed lines, respectively, in Fig.~\textcolor{blue}{\ref{fig:fig4}}. Three different values of $\Gamma$ are shown in the legend. The other parameter values are $M = 0.05$ and $\kappa = 12$. The dashed lines for $J = 0.1$ are shifted along the $\phi$ axis for better visibility. Finite relaxation time of rotational fluctuations of self-propulsion direction associated with the sufficiently large values of $J$ increases the persistence of the disks. Higher persistence helps in MIPS. Hence, the peaks are more prominent for $J = 0.1$. For the remainder of the work, we fix the value of $J$ at $J = 0.01$. In this work we do not study the effect of $J$ further.  
%
%Fig 4

  As $M$ is reduced, the phase boundary moves towards the left along the $\kappa$ axis (see Fig.~\textcolor{blue}{\ref{fig:fig2}}). Hence, in the non-inertial or overdamped limit ($M \rightarrow 0$), relatively softer self-propelled disks phase separate into high density motility induced clusters and a low density phase. At moderate inertia, sufficiently stiffer disks phase separate into MIPS and low density phases; softer disks do not phase separate. At large inertia, $M \geq 0.1$, we do not observe any phase separation in the $\kappa-\Gamma$ plane, and the system remains homogeneous. For a fixed $\kappa$, with increase in  $\Gamma$, the system re-enters the homogeneous phase. In the remainder of the work we investigate the reason behind the above observations.
  
  Sufficiently large density, persistence, and repulsion at the surface of the self-propelled particles are the key ingredients for MIPS. During the collision of two disks, surface interaction of the two disks provide repulsion to block the free movement. Unlike passive disks, due to persistent self-propulsion, two colliding disks form local cluster, hence, provide local nucleation site for cluster growth. Within a persistence time ($\tau_p = 1 / D_r$), if other particles collide with the two-particle-cluster, the cluster grows. For a sufficiently large cluster, disks at the boundary move out of the cluster, when the rotational fluctuations take the self-propulsion direction away form the cluster. A dynamical equilibrium constraint the size of the cluster depending upon the density and the persistence of the disks. During the cluster growth phase, many such clusters are found to merge into a high density cluster, through several mechanism, and the result is a complete phase separation of high density cluster and a low density phase, known as MIPS in literature\cite{ref1}.
  
%
%Fig 5

  In the overdamped limit, due to negligible inertia, the momentum relaxation time scale ($m / \gamma$) becomes very short. Consequently,  the velocity of the particles quickly relaxes towards the self-propulsion direction. Also, the speed or the kinetic energy of the disks is small. On the other hand, in the inertial limit, finite momentum relaxation time sets a natural time delay to the velocity relaxation along the self-propulsion direction. The kinetic energy of the inertial disks are much higher compared to the non-inertial counterpart. Hence, when two disks with large value of inertial parameter $M$ collide, they bounces back and forth, and take finite time before they slow down sufficiently to facilitate motility induced cluster formation. At large inertia, when the momentum relaxation time scale is comparable to the persistence time scale, nucleation sites with two disks form rarely. Hence, we do not observe MIPS at large $M$ (e.g., $M \geq 0.1$) (not shown). However, due to short relaxation time, instead of bouncing back, disks with low value of $M$ slow down quickly. Stiffer walls at large $\kappa$ provide necessary blockage on the free movement of the disks, without any significant deformation in the shape of the disks during collision. Softer disks at small $\kappa$ deform more. Head-on collision of two soft inertial disks still hinder the propulsion of each other. However, when two soft inertial disks approach each other with a large impact parameter ($0 \ll b < \sigma / 2$, see Fig.~\textcolor{blue}{\ref{fig:fig5}}, where we draw a diagram, depicting collision process of two soft SPIDs with large and small $M$), due to large kinetic energy and small stiffness, the shape of the two disks deform, and move apart with minor change in the self-propulsion directions, without forming local two-particle-cluster. Hence, at relatively large inertia $M = 0.05$, we do not observe phase separation for relatively softer inertial disks in the intermediate $\Gamma$ range (e.g., $6 \lesssim \kappa < 12$, $10 \lesssim \Gamma \lesssim 100$). However, in the small inertial limit, due to low kinetic energy of the disks, softer disks can still provide sufficient repulsion to the colliding disks with large impact parameter ($0 \ll b < \sigma / 2$) (see Fig.~\textcolor{blue}{\ref{fig:fig5}}). Hence, MIPS is observed for sufficiently softer disks in the intermediate $\Gamma$ range (e.g, $\kappa \simeq 8$, $10 \leq \Gamma \leq 250$) with low inertial value, $M = 0.005$.
 
\begin{figure}[ht]
  \includegraphics[width=8.5cm, height=6.0cm]{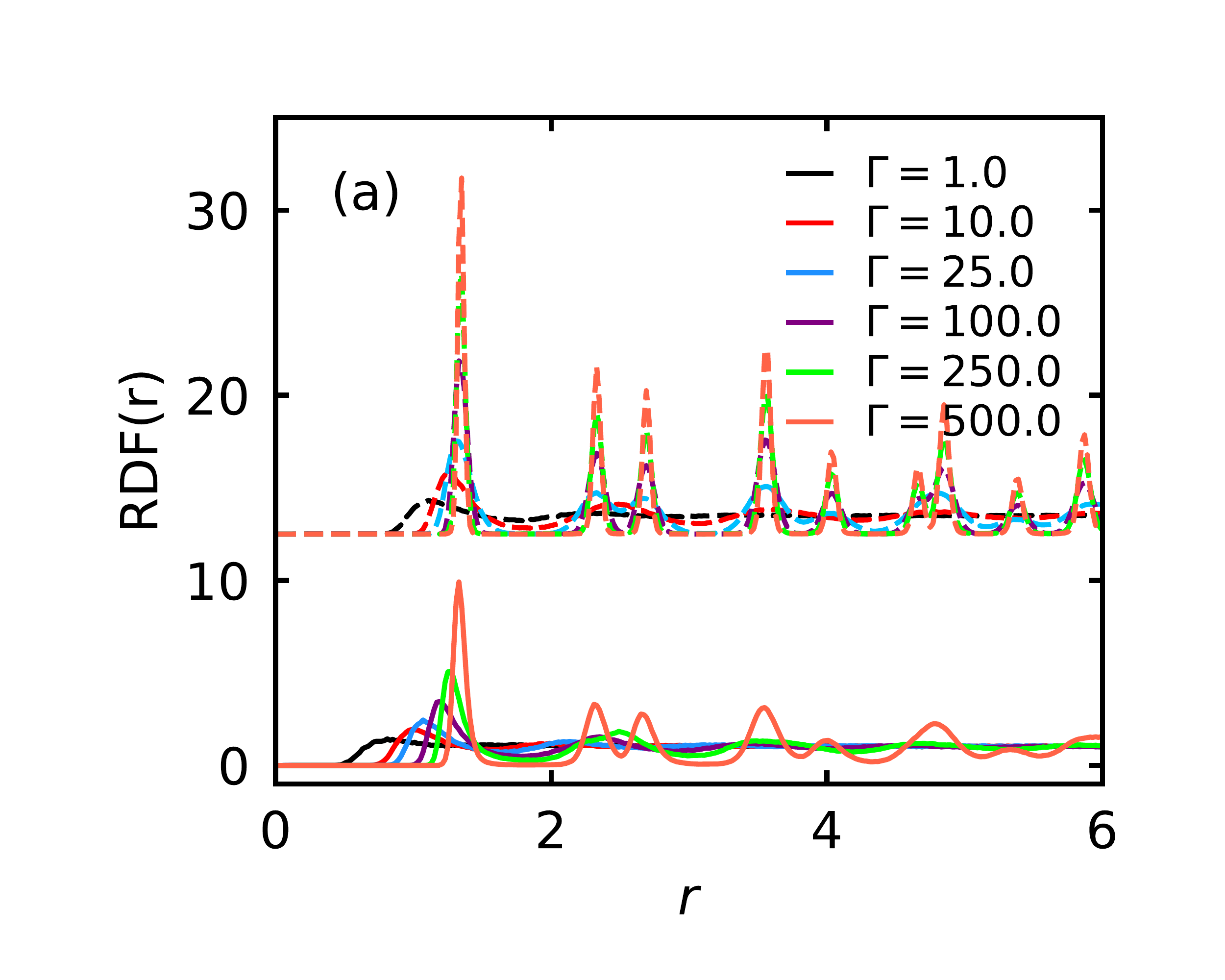}
  \includegraphics[width=8.5cm, height=6.0cm]{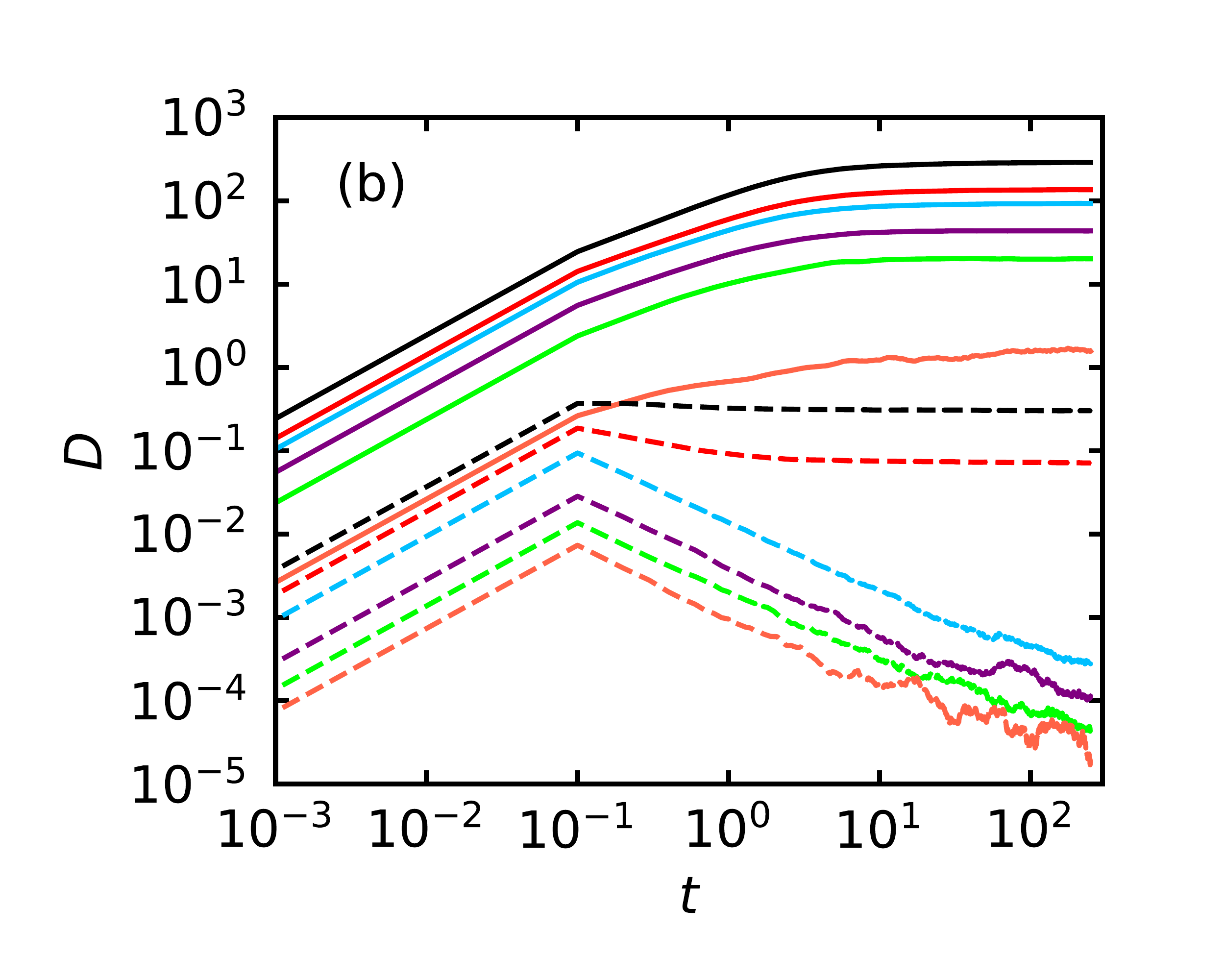}
  \caption{\label{fig:fig6} Plot of radial distribution function (a), and diffusion coefficient with time (b) for various values of $\Gamma$ shown in the legend of (a). Solid lines denote the self-propelled disks with $P = 75$, and the dashed lines denote the passive disks at $P = 0$\cite{ref32}. For better visibility, the dashed lines in (a) are shifted along y-axis by 12.5 units. The values of $M$ and $\kappa$ are $0.05$ and $8$, respectively.}
\end{figure}

  Deep inside the homogeneous region of the phase diagram of our system of inertial self-propelled disks, we plot the radial distribution function in Fig.~\textcolor{blue}{\ref{fig:fig6}(a)} and the diffusion coefficient with time in Fig.~\textcolor{blue}{\ref{fig:fig6}(b)} for various values of $\Gamma$, shown in the legend of RDF plot. The solid lines denote the self-propelled disks with $\textup{P}_e = 75$, and the dashed lines denote the passive disks at $\textup{P}_e = 0$\cite{ref32}. For better visibility, the dashed lines in Fig.~\textcolor{blue}{\ref{fig:fig6}(a)} are shifted by 12.5 units along the y-axis. Inertial parameter is fixed at $M = 0.05$, and the stiffness parameter is fixed at $\kappa = 8$. The other parameter values are considered as before. 
%
%Fig 6
The values of the diffusion coefficient are orders of magnitude lower for the passive disks, compared to their self-propulsive counterpart. Moreover, the diffusion coefficient decreases (beyond the ballistic region) with time for $\Gamma \geq 25$ for the passive disks. The first peak of the radial distribution function of passive disks for $\Gamma = 1$ is at $r \approx 1$, and the location of the first peak increases with increasing $\Gamma$. The second peak of the radial distribution of the passive disks starts to split into two at $\Gamma = 25$. Furthermore, at low stiffness with increasing $\Gamma$, the interaction becomes increasingly longer in range (see Fig.~\textcolor{blue}{\ref{fig:fig1}}). Hence, the system of passive disks transforms from a homogeneous liquid at low $\Gamma$ to an ordered structure at large $\Gamma$, as observed earlier. On the other hand, the system of self-propelled disks remain diffusive for all considered values of $\Gamma$. Due to the persistence of the self-propulsive disks, the location of the first peak is at a distance much less than the diameter of the disks for $\Gamma = 1$. The location of the first peak increases with $\Gamma$, and nearly equals to that of the passive disks at $\Gamma = 500$. The second peak at $\Gamma = 500$ for the self-propelled disks breaks into two. Moreover, there is a density constraint. Hence, at large $\Gamma$, the system of self-propelled disks forms homogeneous ordered state. But, unlike passive disks, the ordered system of self-propelled disks at large $\Gamma$ are diffusive. At larger $\kappa$, the RDF peaks of the self-propelled disks gradually become more prominent (not shown). The location of the first peak at small values of $\Gamma$ is at $r < 1$ (not shown). Therefore, the size of the two-body-clusters are so small that the other disks do not see them because of the constraint of density in the persistence time scale. Hence, the two-body-clusters do not grow much and we do not observe any phase separation at small $\Gamma$ (e.g., at the location of $\kappa = 12$, $\Gamma \leq 10$, for $M = 0.05$ in Fig.~\textcolor{blue}{\ref{fig:fig2}(a)}). At large values of $\Gamma$, the effective diameter of the self-propelled disks (we denote the location of the first peak in the RDF as the effective diameter of the disks) are sufficiently big. Hence, the system re-enters the homogeneous ordered phase (e.g., at the location of $\kappa = 12$, $\Gamma \geq 100$, for $M = 0.05$ in Fig.~\textcolor{blue}{\ref{fig:fig2}(a)}), due to fixed average density of the disks. Furthermore, for softer disks at large $\Gamma$, due to relative long range nature of the interaction, MIPS has not been observed (e.g., $\kappa < 8$, $\Gamma \gtrsim 100$, $M = 0.05$). Similar explanation works for low inertial disks. Hence, we get a parabola shaped phase boundary between homogeneous and the MIPS state for all the inertial parameter values that we incorporate.

%Conclusion
  In this work, we demonstrate that relatively softer disks phase separate due to motility in the inertia-less limit. With increasing inertia, MIPS is observed for self-propelled disks with much stiffer interaction. Moreover, stronger repulsion at larger interaction strength increases the effective diameter of the disks. Hence, the system re-enters the homogeneous ordered state at large interaction strength. Furthermore, long range interaction helps to homogenize the system of self-propelled disks. Hence, long range interactions diminishes MIPS phase space.
  
  It would be interesting to study the structural and dynamical properties of the high and low density phases separately in the soft interaction limit. Janus particles typically has size of $\sigma \sim $ \SI{1}{\micro\metre}, and move at a speed of $v \sim$ \SI{10}{\micro\metre\sec^{-1}} in a background like water at room temperature\cite{ref33}. Hence, Reynolds number ($\textup{Re} = $ inertia forces/viscous forces)\cite{ref34} becomes $\textup{Re} \sim 10^{-3}$, which is an overdamped limit. $\textup{Re}$ of the system can be increased few orders of magnitude either by decreasing the background density, e.g., micron sized particles in activated complex plasma move at high speed in the low density background\cite{ref33, ref35, ref36}, or by increasing the propulsion speed of the motile particles by several mechanisms\citep{ref37, ref38}. Consequently, inertial self-propelled system can be realized. We believe that the present work should be able to bring attention to the novel active matter category of activated complex plasma that could find numerous applications of active materials in the less explored plasma environment. Furthermore, biological materials, e.g., cells, tissues, etc., are much softer in nature compared to artificial self-propelled Janus particles. Our work should help further in understanding the properties of biological active matter.
  
  One of the authors (S. D. K.) would like to thank Devshree Mandal for her help in generating the diagram in Fig.~\textcolor{blue}{\ref{fig:fig5}}. Numerical simulations for this work has been performed on GPU nodes of the ANTYA cluster at Institute for Plasma Research, India.

%-------------------
%References
%-------------------
\bibliography{Reference}

\end{document}